\documentclass[a4paper,11pt]{article}
\usepackage{amssymb, amsmath, amsbsy} 
\usepackage{slashed}
\usepackage{graphicx}
\usepackage{float}
\usepackage{dsfont}
\usepackage[affil-it]{authblk}
\newcommand{\g}{\gamma}
\newcommand{\s}{\slashed}
\usepackage[left=2cm,right=2cm,top=3cm,bottom=3cm]{geometry}
\usepackage{xcolor}
\begin{document} 
\author[1]{J.Lorenzo D\'iaz-Cruz\thanks{\texttt{jldiaz@fcfm.buap.mx}}}
\author[1]{Bryan O. Larios\thanks{\texttt{bryanlarios@gmail.com}}}
\title{\textbf{ Stop Decay with LSP Gravitino in the final state: $\tilde{t}_1\rightarrow\,\widetilde{G}\,W\,b$}}
\affil{Facultad de Ciencias F\'isico - Matem\'aticas, BUAP \protect\\Apdo. Postal 1364, C.P. 72000, Puebla, Pue. M\'exico} 

\maketitle
\begin{abstract}
In MSSM scenarios  where the gravitino is the lightest supersymmetric particle (LSP), 
and therefore a viable dark matter candidate, the stop $\tilde{t}_1$ could be the next-to-lightest 
superpartner (NLSP). For a mass spectrum satisfying:
$m_{\widetilde{G}}+m_t>m_{\tilde{t}_1}>m_{\widetilde{G}}+m_b+m_W$, the stop decay is
dominated by the 3-body mode $\tilde{t}_1\rightarrow b\,W\,\tilde{G}$. 
We calculate the stop life-time, including the full contributions from top, sbottom and chargino 
as intermediate states. We also evaluate the stop  lifetime for the case when the gravitino can be approximated by 
the goldstino state. Our analytical results are conveniently expressed using an expansion in 
terms of the intermediate state mass, which helps  to identify the massless limit. 
 In the region of low gravitino mass ($m_{\widetilde{G}}\ll m_{\tilde{t}_1}$) 
the results obtained using the gravitino and goldstino cases turns out to be similar, as expected.
However for higher gravitino masses $m_{\widetilde{G}} \lesssim m_{\tilde{t}_1}$ 
the results for the lifetime  could show a difference of O(100)\%.
\end{abstract}   

 \unitlength=1mm

\newpage
\section{Introduction}\label{introduction}
The properties of Supersymmetric theories, both in the ultraviolet or the infrared domain have had a great 
impact in distinct  domains of particle physics, including model building, phenomenology, 
cosmology and formal quantum field theory \cite{SUSYPrimer}.  
In particular, Supersymmetric extensions of the Standard Model can include a discrete symmetry,  $R$ parity, 
that guarantees the stability of the lightest supersymmetric particle (LSP) \cite{EHNOS}, 
which allows the LSP to be a good candidate for dark matter 
(DM). Candidates for the LSP in the minimal supersymmetric
extension of the Standard Model (MSSM) include sneutrinos, the
lightest neutralino $\chi^0_1$ and the gravitino ${\widetilde G}$. 
Most studies  has focused on the neutralino LSP \cite{vcmssm}, while 
scenarios with the sneutrino LSP seem more constrained ~\cite{FOS94}. 

Scenarios with gravitino LSP as DM candidate have also been considered \cite{FengGDM,gdm,FengGDM2}. 
In such scenarios, the nature of the next-to-lightest supersymmetric particle (NLSP) 
determines its phenomenology~\cite{steffan, Bench3}.

Possible candidates for NLSP include the lightest neutralino~\cite{Neutralino,Neutralino2}, 
the chargino~\cite{chargino}, the lightest charged slepton~\cite{cslepton},  or  
the sneutrino~\cite{FengGDM3,KKKM,lcovi,lvelazco}.  
The NLSP could have a long lifetime, due to the weakness of the gravitational interactions,  
and this leads to scenarios with a metastable charged sparticle that  could have dramatic signatures at
colliders~\cite{Are,Nojiri} and it could also affect the Big Bang nucleosynthesis (BBN)~\cite{CEFO,KKM,kohri}.

Squark species could also be the NLSP, and in such case natural candidates for NLSP 
could be the sbottom~\cite{ER,BDD,stopco} or the lightest stop $\tilde{t}_1$.  
There are several experimental and cosmological constraints for
the  scenarios with a gravitino LSP and a stop NLSP that were discussed in \cite{jlorenzo}. 
It turns out that the lifetime of the stop ${\tilde t_1}$ could be (very) long, in which case the
relevant collider limits are those on (apparently) stable charged particles. For instance the limits available from the 
Tevatron collider imply that $m_{\tilde t_1} > 220$~GeV~\cite{Tevatron}~\footnote[1]{The LHC will probably be sensitive to a
metastable ${\tilde t_1}$ that is an order of magnitude heavier.}.
Thus, knowing in a precise way the stop lifetime is one of the most important issues in this scenario, and this is 
precisely the goal of our work. In this paper we present a detailed calculation of the stop lifetime, for the kinematical 
region where the 3-body mode $\tilde{t}_1\rightarrow\,\widetilde{G}\,W\,b$ dominates\footnote[2]{Our calculation of stop 
lifetime improves the one presented in \cite{jlorenzo} where an approximation was used for the chargino-mediated 
contribution that neglected a subdominant term in the expression for the vertex $\chi_i^+\,\widetilde{G}\,W$.}.
Besides calculating the amplitude using the full wave function for the gravitino, we have also calculated the 
3-body decay width (and lifetime) using the
gravitino-goldstino equivalence theorem~\cite{goldstino}.  
It should be  mentioned that this scenario is not viable within  the Constrained Minimal Supersymmetric Standard Model (CMSSM). However there are regions of parameter space within the Non-Universal Higgs Masses model (NUHM) that pass all collider and cosmological constraints (relic density, nucleosynthesis, CMB mainly)~\cite{covi}.

 

The organization of our paper goes as follows, we begin Section \ref{stoplifetime} by giving some formulae for the stop mass. 
In Section \ref{gravitinoamplitud} we compute the squared amplitudes for the stop decay with gravitino in the final 
state ($\tilde{t}_1\rightarrow\,\widetilde{G}\,W\,b$) including the chargino, sbotom and top mediated states. After carefully 
analyzing the results for the squared amplitude, we have identified a convenient expansion in terms of powers of the 
intermediate particle mass, which only needs terms of order $O(m_i),\, O(m_i m_j)$. It is our hope that such expansion 
could help in order to relate the calculation of the massive and massless cases. In future work we plan to reevaluate 
this decay using the helicity formalism suited for the spin-$\frac{3}{2}$ case. In Section \ref{goldstinoamplitud} we compute the 
squared amplitudes for the stop decay considering the gravitino-goldstino high energy equivalence theorem that allow us 
approximate the gravitino as the derivative of the goldstino. 
We present in Section \ref{numericalsection} our numerical results, showing some plots where we reproduce  the stop lifetime for 
the approximate amplitude considered in \cite{jlorenzo}, and compare it with our complete calculation, we also compare 
these results with goldstino approximation. Conclusions are included in Section \ref{conclusions}, finally  
 all the analytic full results for the squared amplitudes are left in Appendices \ref{apendixA},\ref{appendixB}.  

\section{The Stop Lifetime within the MSSM}\label{stoplifetime}
We start by giving some  relevant formulae for the input parameters that appear in the Feynman rules of the gravitino
within the MSSM.
The (2x2) stop mass matrix can be written as:
\begin{equation}
\widetilde{M}_{\tilde{t}}^2 =
 \begin{pmatrix}
  M_{LL}^2  & M_{LR}^2 \\
 M_{LR}^{2\,\dag} & M_{RR}^2\\
 \end{pmatrix},
 \label{eq:01}
 \end{equation}
where the entries take the form:
\begin{align} \nonumber
M_{LL}^2 &= M_{L}^2+m_t^2+\frac{1}{6}\cos2\beta \,(4m_W^2-m_Z^2),\\
M_{RR}^2 &= M_{R}^2+m_t^2+\frac{2}{3}\cos2\beta\sin^2\theta_W\, m_Z^2,\\ \nonumber
M_{LR}^2 &= -m_t (A_t + \mu \, \cot \beta) \equiv - m_t X_t\,.
\end{align}
The corresponding mass eigenvalues are given by:
\begin{equation}
m^2_{\tilde{t}_1}=m_t^2 + \frac{1}{2}(M_{L}^2+  M_{R}^2)+
\frac{1}{4}m^2_Z \cos 2\beta-\frac{\Delta}{2}   ,   
\end{equation}
and
\begin{equation}
m^2_{\tilde{t}_2}= m^2_t + \frac{1}{2}(M_{L}^2+  M_{R}^2)+
\frac{1}{4}m^2_Z \cos 2\beta+\frac{\Delta}{2}    ,            
\end{equation}
where
$\Delta^2= \left( M_{L}^2 -  M_{R}^2 + \frac{1}{6} \cos 2\beta (8
m^2_W-5m^2_Z) \right)^2 + 4\, m_t^2 |A_t + \mu \cot \beta |^2$.
The mixing angle $\theta_{\tilde{t}}$ appears in the   mixing matrix that relate the 
weak basis $(\tilde{t}_L,\tilde{t}_R)$ and the mass
eigenstates  $(\tilde{t}_1,\tilde{t}_2)$, and it is given by
$\tan \theta_{\tilde{t}}= \frac{(m^2_{\tilde{t}_1}-M^2_{LL})}{|M^2_{LR}|}$.
From these expressions it is clear that
in order to obtain a very light stop one needs to have a
very large value for the trilinear soft supersymmetry-breaking parameter~\cite{stopco,ben}. 
It turns out that  such scenario helps to obtain a Higgs mass value in agreement with the mass 
measured at LHC (125-126 GeV)  in a consistent way within the MSSM. 

Following Ref.~\cite{moroi}, we derived the  expressions for all the relevant interactions vertices that  
appear in the amplitudes for the decay width ($\tilde{t}_1\rightarrow\,\widetilde{G}\,W\,b$), 
whose Feynman graphs are shown in Figures~[\ref{fig:topgravitino}-\ref{fig:charginogravitino}]. 
We shall need the following vertices:
\begin{align}
V_1(\tilde{t}_1\,t\,\widetilde{G})& = -\frac{1}{\sqrt{2}M}(\g^{\nu}\g^{\mu}p_{\nu})(\cos\theta_{\tilde{t}}P_R+\sin\theta_{\tilde{t}}P_L),\label{eq:02}\\ 
V_2(t\,b\,W)&=\frac{ig_2}{\sqrt{2}}\g_{\rho}P_L,\label{eq:03}\\
V_3(\tilde{t}_1\,W\,\tilde{b}_i)&=-\frac{ig_2\kappa_i}{\sqrt{2}}(p+q_1)_{\mu}\,,\label{eq:202}\\ 
V_4(\tilde{b}_i\,b\,\widetilde{G}) & = -\frac{1}{\sqrt{2}M}(\g^{\nu}\g^{\mu}q_{2\nu})(a_iP_R+b_iP_L),\label{eq:303}\\ 
V_5(\tilde{t}_1\,b\,\chi_i^+) &=-i(S_i+P_i\g_5),\label{eq:04}\\ 
V_6(\chi_i^+\,W\,\widetilde{G})&=-\frac{1}{\sqrt{2}M}\Big(-\frac{1}{4}\s{p}\g^{\rho}\g^{\mu}(V_{1i}P_R-U_{i1}P_L)\label{eq:05}\\
&\quad-m_{W}\g^{\nu}\g^{\mu}(V_{i2}\sin{\beta}P_R+U_{i2}\cos\beta P_L)\Big),
\nonumber
\end{align}
where $\tilde{t}_1$ denotes the lightest stop, while $t$ is the top quark and $\widetilde{G}$ denotes the gravitino. 
With $b$ we denote the bottom quark, while $W$ is the gauge boson, $\chi_i^+$ denotes the chargino and $\tilde{b}_i$ is de sbottom. 
With $P_R$ and $P_L$ corresponding to  the left and right projectors, $a_i\,b_i,\,S_i,\,P_i$ are defined in 
Appendices \ref{apendixA}, \ref{appendixB}, as well the mixing matrices $V_{1i}$, $U_{1i}$ that diagonalize the chargino factor. 

For the case when the gravitino approximates to the goldstino state,  the interaction vertices that will appear  in the amplitudes for the decay width ($\tilde{t}_1\rightarrow\,G\,W\,b$) are the following:
\begin{align}
\widetilde{V}_1(\tilde{t}_1\,t\,G)& = \left(\frac{m_{t}^2-m_{\tilde{t}_1}^2}{2\sqrt{3}Mm_{\widetilde{G}}}\right)(\cos\theta_{\tilde{t}}P_R+\sin\theta_{\tilde{t}}P_L),\label{eq:06}\\ 
\widetilde{V}_4(\tilde{b}_i\,b\,G) & = \left(\frac{m_b^2-m_{\tilde{b}_i}^2}{2\sqrt{3}Mm_{\widetilde{G}}}\right)(a_iP_R+b_iP_L),\label{eq:07}\\ 
 \widetilde{V}_6(\chi_i^+\,W\,G)&=-\frac{m_{\chi_i^+}}{\sqrt{6}Mm_{\widetilde{G}}}[\s{p}\g^{\rho}(V_{1i}P_R-U_{i1}P_L)],\label{eq:08}
\end{align}
whereas the vertices $V_2(t\,b\,W),\,V_3(\tilde{t}_1\,W\,\tilde{b}_i)$ and $V_5(\tilde{t}_1\,b\,\chi_i^+)$ remain the same as in the gravitino case.

\subsection{The Amplitude for  $\tilde{t}_1\rightarrow\,\widetilde{G}\,W\,b$}\label{gravitinoamplitud}
The decay lifetime of the stop was calculated in Ref.\cite{jlorenzo}, where the chargino contribution was approximated by including 
only the dominant term. Here we shall calculate the full amplitude and determine the importance of the neglected term for 
the numerical calculation of the stop lifetime.
In what follows we need to consider the Feynman diagrams shown in Figures
[\ref{fig:topgravitino},\ref{fig:sbottomgravitino},\ref{fig:charginogravitino}], 
which contribute to the decay amplitude for  $\tilde{t}_1(p)\rightarrow\,\widetilde{G}(p_1)\,W(k)\,b(p_2)$, 
with the momenta assignment shown in parenthesis.
\\

\begin{minipage}{\linewidth}
      \centering
      \begin{minipage}{0.3\linewidth}
          \begin{figure}[H]
\centering
\begin{picture}(60,30)
\put(10,0){\includegraphics[scale=0.5]{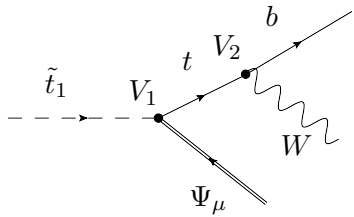}}
\put(15,15){$\tilde{t}_1$}
\put(33,18){$t$}
\put(44,23.8){$b$}
\put(46,6.7){$W$}
\put(34,0){$\Psi_{\mu}$}
\put(26,14){$V_{1}$}
\put(37,20){$V_{2}$}
\end{picture}
        \caption{Top mediated diagram}
        \label{fig:topgravitino}
\end{figure}
      \end{minipage}
      \hspace{0.05\linewidth}
      \begin{minipage}{0.3\linewidth}
 \begin{figure}[H]
\centering
\begin{picture}(60,30)
\put(10,0){\includegraphics[scale=0.5]{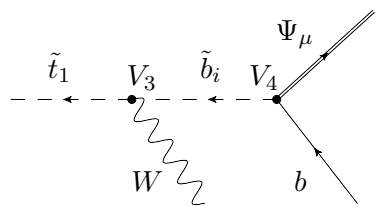}}
\put(15,17){$\tilde{t}_1$}
\put(35,17){$\tilde{b}_{i}$}
\put(41.5,16){$V_{4}$}
\put(45,22){$\Psi_{\mu}$}
\put(47.4,2){$b$}
\put(26,2){$W$}
\put(25.4,16){$V_{3}$}
\end{picture}
        \caption{Sbottom mediated diagram}
        \label{fig:sbottomgravitino}
\end{figure}
\end{minipage}
\hspace{0.1\linewidth}
\begin{minipage}{0.6\linewidth}
\begin{figure}[H]
\centering
\begin{picture}(60,30)
\put(10,0){\includegraphics[scale=0.5]{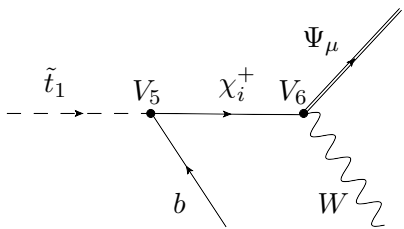}}
\put(15,18){$\tilde{t}_1$}
\put(38,18){$\chi_i^+$}
\put(45.8,17){$V_{6}$}
\put(49,24){$\Psi_{\mu}$}
\put(51,2){$W$}
\put(32,2){$b$}
\put(26.6,17){$V_{5}$}
\end{picture}
        \caption{Chargino mediated diagram}
        \label{fig:charginogravitino}
\end{figure}
      \end{minipage}
      \vspace{0.05\linewidth}
      \end{minipage}
The total  amplitude is given by:
\begin{equation}\label{eq:09}
\mathcal{M}=\mathcal{M}_t+\mathcal{M}_{\tilde{b}_i}+\mathcal{M}_{\chi_i^+}^C\,,
\end{equation}
where $\mathcal{M}_t,\,\mathcal{M}_{\tilde{b}_i},\,\mathcal{M}_{\chi_i^+}^C$ denotes the amplitudes for top, sbottom and chargino 
mediate diagrams, respectively.  In the calculation of Ref.~\cite{jlorenzo}, the chargino-mediated diagram included only part of the vertex $V_6(\chi_i^+\,W\,\widetilde{G})$.
Here, in order to keep control of the vertex $V_6$ and therefore $\mathcal{M}_{\chi_i^+}^c$,  we shall split $\mathcal{M}_{\chi_i^+}^c$  into two terms as follows
\begin{equation}\label{eq:10}
\mathcal{M}_{\chi_i^+}^c=\mathcal{M}_{\chi_{i}^+}^0+\mathcal{\widetilde{M}}_{\chi_{i}^+}\,,
\end{equation}
where $\mathcal{M}_{\chi_{i}^+}^0$  denotes  the amplitude considered in Ref.~\cite{jlorenzo}, which only includes the second term of [\ref{eq:05}] (with two gamma matrices), while $\mathcal{\widetilde{M}}_{\chi_i^+}$  includes the first term  (with 3 gamma matrices).
Then, the averaged squared amplitude [\ref{eq:09}] becomes 
\begin{align}\label{eq:11} \nonumber
\mid\overline{\mathcal{M}}\mid^2&=\mid\mathcal{M}_t\mid^2+\mid\mathcal{M}_{\tilde{b}_i}\mid^2+\mid\mathcal{M}_{\chi_i^+}^0\mid^2+\mid\mathcal{\widetilde{M}}_{\chi_i^+}\mid^2+2\,\textbf{Re}\Big(\mathcal{M}_{\chi_i^+}^{0\,\dagger}\mathcal{\widetilde{M}}_{\chi_i^+}+\mathcal{M}_{t}^{\dagger}\mathcal{M}_{\tilde{b}_i}
\\
&\quad+\mathcal{M}_t^{\dagger}\mathcal{M}_{\chi_i^+}^0+\mathcal{M}_t^{\dagger}\mathcal{\widetilde{M}}_{\chi_i^+}+\mathcal{M}_{\tilde{b}_i}^{\dagger}\mathcal{M}_{\chi_i^+}^0+\mathcal{M}_{\tilde{b}_i}^{\dagger}\mathcal{\widetilde{M}}_{\chi_i^+}\Big).
\end{align}
From the inclusion of  the vertices $V_i$  from each graph, we can build each  amplitudes, as follows:
\begin{align}
\mathcal{M}_t &=C_t P_t(q_1)\,\overline{\Psi}_{\mu}p^{\mu}(A_{t}+B_{t}\gamma_5)(\slashed{q_1}+m_t)\gamma^{\rho}\epsilon_{\rho}(k)P_L u(p_2) \label{eq:12},\\ 
\mathcal{M}_{\tilde{b}_i} &=C_{\tilde{b}_i} P_{\tilde{b}_i}(q_2)\,\overline{\Psi}_{\mu}q_2^{\mu}(a_i P_l+b_i P_R)p^{\rho}\epsilon_{\rho}(k)P_L u(p_2)\label{eq:13}, \\ 
\mathcal{M}_{\chi_{i}^{+}}^0 &=C_{\chi_{i}^{+}}^0 P_{\chi_{i}^{+}}(q_3)\,\overline{\Psi}_{\mu}\gamma^{\rho}\epsilon_{\rho}(k)\gamma^{\mu}(V_i+\Lambda_i\g_5)(\s{q_3}+m_{\chi})(S_i+P_i\g_5) u(p_2)\label{eq:56:chargino:oldamplitude},\\
\mathcal{\widetilde{M}}_{\chi_i^+}&=C_{\chi_i^+}P_{\chi_i^+}\overline{\Psi}_{\mu}\s{p}\g^{\rho}\g^{\mu}(T_i+Q_i\g_5)\epsilon_{\rho}(k)(\s{q}_3+m_{\chi})(S_i+P_i\g_5)u(p_2).
\end{align}
Where $C_t=\frac{g_2}{2M}$, $C_{\tilde{b}_i}=\frac{g_2 \kappa_i}{M}$, $C_{\chi_{i}^{+}}^0=\frac{m_W}{M}$ and $C_{\chi_i^+}=\frac{1}{8M}$. We have defined $q_1\equiv p-p_1,\, q_2\equiv p-k$ and $q_3\equiv p-p_2$, and $\epsilon_{\rho}(k)$ denotes the $W$ polarization vector. Expressions for $A_{\tilde{t}},B_{\tilde{t}},a_i,b_i,\kappa_i,V_i,A_i,S_i$ and $P_i$ are presented in the Appendices \ref{apendixA},\ref{appendixB}. 
Then, after performing the evaluation of each expression, we find convenient to express each squared amplitude, as follows:
\begin{equation}\label{eq:14}
\mid\mathcal{M}_{\psi_a} \mid^2=C_{\psi_a}^2\mid P_{\psi_a}(q_a)\mid^2W_{\psi_a\psi_a},
\end{equation}
where $\psi_a=(t,\tilde{b}_j,\chi_k^+)$.
The functions $P_{\psi_a}(q_a)$ correspond to the propagators factors, thus for the chargino $\psi_a=\chi_i^+$, we have
\begin{equation}\label{eq:75:propagators}
P_{\chi_i^+}(q_3)=\frac{1}{q_3^2-m_{\chi_i^+}^2+i\epsilon}.
\end{equation}
Similar expressions hold for the sbottom and the top contributions, $P_{\tilde{b}}(q_2)$ and $P_t(q_1)$ respectively.
The terms $W_{\psi_a\psi_a}$ include the traces involved in each squared amplitudes
\begin{align}
W_{tt} &= \mathbf{Tr}\Big[M_{\rho\sigma}D_{\mu\nu}p^{\mu}p^{\nu}(A_{\tilde{t}}+B_{\tilde{t}}\g_5)(\s{q}_1+m_t)\g^{\rho} \nonumber\\
&\quad P_L\s{p}_2P_R\g^{\sigma}(\s{q}_1+m_t)(A_{\tilde{t}}-B_{\tilde{t}}\g_5)\Big] \label{topWfunction},\\
W_{\tilde{b}_i\tilde{b}_i}&=\mathbf{Tr}\Big[p^{\rho}p^{\sigma}M_{\rho\sigma}D_{\mu\nu}q_2^{\mu}q_2^{\nu}(R_i+Z_i\g_5)\s{p_2}(R_j-Z_j\g_5)\Big]\label{sbottomWfunction},\\
W_{\chi_{i}^{+}\chi_{i}^{+}}^0&=\mathbf{Tr}\Big[M_{\rho\sigma}D^{\rho\sigma}(V_i+\Lambda_i\g_5)(\s{q_3}+m_{\chi})(S_i+P_i\g_5)\s{p}_2\\ \nonumber
&\quad (S_j-P_j\g_5)(\s{q_3}+m_{\chi})(V_j-\Lambda_j\g_5)\Big]\label{chargino0Wfunction},\\
W_{\chi_i^+\chi_i^+} &= \mathbf{Tr}\Big[M_{\rho\sigma}D_{\mu\nu}\s{p}\g^{\rho}\g^{\mu}(T_i+Q_i\g_5)(\s{q}_3+m_{\chi})(S_i+P_i\g_5)\s{p}_2(S_j-P_j\g_5)\\ \nonumber
&\quad(\s{q}_3+m_{\chi})(T_j-Q_j\g_5)\g^{\nu}\g^{\sigma}\s{p}\Big]\nonumber.
\end{align}
For  simplicity, we have written the completeness relations for the gravitino field  and the vector polarization sum of the boson W  as follows:
\begin{align}\label{eq:completeness}
\sum_{\lambda =1}^{3}\epsilon_{\rho}(\vec{k},\lambda)\epsilon_{\sigma}^*(\vec{k},\lambda)&=-g_{\rho\sigma}+\frac{k_{\rho}k_{\sigma}}{m_W^2}=M_{\rho\sigma}\\\sum_{\tilde{\lambda}=1}^3\Psi_{\mu}(\vec{p}_1,\tilde{\lambda})\overline{\Psi}_{\nu}(\vec{p}_1,\tilde{\lambda}) &=-(\s{p}_1+m_{\tilde{G}})\times
\left\{\left(g_{\mu\nu}-\frac{p_{\mu}p_{\nu}}{m_{\tilde{G}}^2}\right)\right. \\
&\quad \left.-\frac{1}{3}\Bigg(g_{\mu\sigma}-\frac{p_{\mu}p_{\sigma}}{m_{\tilde{G}}^2}\Bigg)\Bigg(g_{\nu\lambda}-\frac{p_{\nu}p_{\lambda}}{m_{\tilde{G}}^2}\Bigg)\g^{\sigma}\g^{\lambda} \right\}=D_{\mu\nu}.
\end{align}
The functions $W_{\psi_a\psi_a}$  depend on the scalar products of the momenta $p, p_1, p_2, k, q_1, q_2$ and $q_3$. After carefully analyzing the resulting traces (handed with FeynCalc\footnote{Progress in automatic calculation of MSSM processes with gravitino have appeared recently \cite{spanos}, some of our results have been checked by the authors of Ref.~\cite{spanos2} and they found agreement in the results (private communications).} \cite{FC1,FC2}) we find that these functions can be written as powers of the intermediate state masses, as follows: 
\begin{equation}\label{eq:72:Wfunctions}
W_{\psi_a\psi_a}=\text{w}_{1\psi_a\psi_a}+m_{\psi_a}\text{w}_{2\psi_a\psi_a}+m_{\psi_a}^2\text{w}_{3\psi_a\psi_a}.
\end{equation}
Full expressions for each function $\text{w}_{i\psi_a\psi_a}$ $\forall\,i=1,2,3$  are included in Appendix \ref{apendixA}. 
Furthermore, we also find that the interference terms can be written in a similar form, namely:
\begin{equation}\label{eq:73:interferences}
\mathcal{M}_{\psi_a}^{\dagger}\mathcal{M}_{\psi_b}=C_{\psi_a}C_{\psi_b}P^*_{\psi_a}(q_a)P_{\psi_b}(q_b)W_{\psi_a\psi_b}.
\end{equation}
Again, as in the previous case, the function 
$W_{\psi_a\psi_b}$ include the traces appearing in the interferences, specifically  we have 
\begin{align}
\widetilde{W}_{\chi_{i^+}\chi_{i^+}}&=\mathbf{Tr}\Big[M_{\rho\sigma}D_{\mu\nu}\s{p}\g^{\rho}\g^{\mu}(T_i+Q_i\g_5)(\s{q}_3+m_{\chi})\s{p}_2(S_i-P_i\g_5)(S_j-P_j\g_5)\\ \nonumber
&\quad(\s{q}_3+m_{\chi})(V_j-\Lambda_j\g_5)\g^{\nu}g^{\sigma}\Big],\nonumber\\
W_{t\tilde{b}_i}&=\mathbf{Tr}\Big[M_{\rho\sigma}p^{\rho}\s{p}_2P_R\g^{\sigma}(\s{q}_1+m_t)(A_{\tilde{t}}-B_{\tilde{t}}\g_5)p^{\mu}D_{\mu\nu}q_2^{\nu}(R_{i}+Z_{i}\g_5)\Big], \\ 
W_{t\chi_i^+} &=\mathbf{Tr}\Big[M_{\rho\sigma}\s{p}_2P_R\g^{\sigma}(\s{q}_1+m_t)(A_{\tilde{t}}-B_{\tilde{t}}\g_5)p_{\mu}D^{\mu\rho}(\Lambda_i+V_i\g_5)(\s{q}_3+m_{\chi})\\ \nonumber
&\quad(S_i+P_i\g_5)\Big],\\
\widetilde{W}_{t\chi_{i}^+}&=\mathbf{Tr}\Big[M_{\rho\sigma}D_{\mu\nu}\s{p}\g^{\rho}\g^{\mu}p^{\nu}(T_i+Q_i\g_5)(\s{q}_3+m_{\chi})(S_i+P_i\g_5)\s{p}_2P_R\g^{\sigma}\\\nonumber
&\quad(\s{q}_1+m_t)(A_{\tilde{t}}-B_{\tilde{t}}\g_5)\Big],\\ 
W_{\chi_i^+\tilde{b}_i}&=\mathbf{Tr}\Big[M_{\rho\sigma}p^{\rho}(p_{\nu}-k_{\nu})\s{p}_2(S_i-P_i\g_5)(\s{q}_3+m_{\chi})(\Lambda_i-V_i\g_5)\\ \nonumber
&\quad D^{\nu\sigma}(R_j+Z_j\g_5)\Big],\\
\widetilde{W}_{\chi_i^+\tilde{b}_i} &=\mathbf{Tr}\Big[M_{\rho\sigma}D_{\mu\nu}(p^{\nu}-k^{\nu})(R_i+Z_i\g_5)\s{p}_2(S_i-P_i\g_5)(\s{q}_3+m_{\chi})\\\nonumber
&\quad(T_i-Q_i\g_5)\g^{\mu}\g^{\rho}\s{p}p^{\sigma}\Big].
\end{align}
It turns out that the functions $W_{\psi_a\psi_b}$ can be expressed also in powers of the intermediate masses:
\begin{equation}\label{eq:74:Wfunct:int}
W_{\psi_a\psi_b}=\text{w}_{1\psi_a\psi_b}+m_{\psi_a}(\text{w}_{2\psi_a\psi_b}+m_{\psi_b}\text{w}_{3\psi_a\psi_b})+m_{\psi_b}\text{w}_{4\psi_a\psi_b}.
\end{equation}
The  $\text{w}_{j\psi_a\psi_b}$ $\forall\,j=1,2,3,4$ are as the $\text{w}_{i\psi_a\psi_a}$ 4-momentum's scalar products functions completely determined by the kinematics of our decay.
We consider that [\ref{eq:72:Wfunctions}] and [\ref{eq:74:Wfunct:int}] are an useful way to present our results as well an easy manner to compute complicated and messy traces. 
Then the decay width can be obtained after integration of the 3-body phase-space  
\begin{equation}\label{eq:76:widthdecay}
\frac{d\Gamma}{dx\,dy}=\frac{m_{\tilde{t}_1}^2}{256\,\pi^3}\mid\overline{\mathcal{M}} \mid^2.
\end{equation}
The  variables $x$ and $y$ are defined as $x =2 \frac{E_{\widetilde{G}}}{m_{\tilde{t}_1}}$ and $y =2 \frac{E_{W}}{m_{\tilde{t}_1}}$. Numerical results for the lifetime $\tau=\frac{1}{\Gamma}$ will be presented and discussed in Section \ref{numericalsection}.
   \subsection{The Amplitudes $\tilde{t}_1\rightarrow\,G\,W\,b$ with the goldstino approximation}\label{goldstinoamplitud}
    In this section we shall present the calculation of the stop decay using the gravitino-goldstino high energy equivalence theorem \cite{goldstino}. In the high energy limit ($m_{\widetilde{G}}\ll m_{\tilde{t_1}}$) we could consider the gravitino field (spin $\frac{3}{2}$  particle) as the derivative of the goldstino field (spin $\left(\frac{1}{2}\right)$  particle).
  We shall consider in this section the same Feynman diagrams  Figures [\ref{fig:topgravitino},\ref{fig:sbottomgravitino},\ref{fig:charginogravitino}] that 
  we used in Section \ref{gravitinoamplitud},  but with the proviso that  the gravitino field shall be described by the goldstino fields. Making the replacement $\Psi_{\widetilde{G}}\to i\sqrt{\frac{2}{3}}\frac{1}{m_{\widetilde{G}}}\,\partial_{\mu}\Psi$
  in the gravitino interaction lagrangian, one obtain the effective interaction lagrangian for the goldstino as is show in \cite{moroi}. 
The averaged squared amplitude for the Goldstino is then written as
\begin{align}\label{eq:goldstinoamplitude}
\mid\overline{\mathcal{M}^G}\mid^2&=\mid\mathcal{M}_t^G\mid^2+\mid\mathcal{M}_{\tilde{b}_i}^G\mid^2+\mid\mathcal{M}_{\chi_i^+}^G\mid^2\\\nonumber
&\quad+2\,\textbf{Re}(\mathcal{M}_{t}^{G\,\dagger}\mathcal{M}^G_{\tilde{b}_i}+\mathcal{M}_t^{G\,\dagger}\mathcal{M}_{\chi_i^+}^G+\mathcal{M}_{\tilde{b}_i}^{G\,\dagger}\mathcal{M}_{\chi_i^+}^G). 
\end{align}
As in the previous Section \ref{gravitinoamplitud},  we can build  the amplitudes from the inclusion of all the vertices into the expressions from each graph, namely:
  \begin{align}
  \mathcal{M}_t^G &= \widetilde{C}_tP_t(q_1)\overline{\Psi}(A_{\tilde{t}}+B_{\tilde{t}}\g_5)(\s{q}_1+m_t)\g^{\rho}P_L\epsilon_{\rho}(k)u(p_2)\label{eq:113:goldstino:top:amp},\\
  \mathcal{M}_{\tilde{b}_i}^G&=\widetilde{C}_{\tilde{b}_i}P_{\tilde{b}_i}(q_2)\overline{\Psi}(R_i+Z_i\g^5)u(p_2)p^{\sigma}\epsilon_{\sigma}(k)\label{eq:114:goldstino:sbotom:amp},\\
  \mathcal{M}_{\chi_i^+}^G &= \widetilde{C}_{\chi_i^+}P_{\chi_i^+}(q_3)\s{p}\g^{\rho}(T_i+Q_i\g_5)\overline{\Psi}\epsilon_{\rho}(k)(\s{q}_3+m_{\chi})(S_i+P_i\g_5)u(p_2)\label{eq:115:goldstino:chargino:amp}.
  \end{align}
Where the superindex  ``G'' that appears in the amplitudes [\ref{eq:113:goldstino:top:amp}-\ref{eq:115:goldstino:chargino:amp}] refers to the goldstino amplitudes. The constants appearing in front of each amplitudes are: $\widetilde{C}_t=-g_2\left(\frac{m_{t}^2-m_{\tilde{t}_1}^2}{4\sqrt{6}Mm_{\widetilde{G}}}\right)$, $\widetilde{C}_{\tilde{b}_i}=g_2\kappa_i\left(\frac{m_b^2-m_{\tilde{b}_i}^2}{4\sqrt{6}Mm_{\widetilde{G}}}\right)$ and $\widetilde{C}_{\chi_i^+}=-\frac{m_{\chi_i^+}}{\sqrt{6}Mm_{\widetilde{G}}}$. We obtain similar expressions to [\ref{eq:71:amplitudes}] for the squared amplitudes of the goldstino case, namely:
\begin{equation}\label{eq:amplitudesgoldstino}
\mid\mathcal{M}_{\psi_a}^G \mid^2=\widetilde{C}_{\psi_a}^2\mid P_{\psi_a}(q_a)\mid^2W_{\psi_a\psi_a}^G,
\end{equation}
where the  function $W_{\psi_a\psi_a}^G$ includes traces corresponding to the goldstino squared amplitudes,  which are given as follows: 
\begin{align}
W_{tt}^G &=\mathbf{Tr}\Big[(\s{p}_1+m_{\tilde{G}})(A_{\tilde{t}}+B_{\tilde{t}}\g_5)(\s{q}_1+m_t)\g^{\rho}P_LM_{\rho\sigma}\s{p}_2 
\\\nonumber
&\quad P_R\g^{\sigma}(\s{q}_1+m_t)(A_{\tilde{t}}-B_{\tilde{t}}\g_5)\Big],\\
W^G_{\tilde{b}_i\tilde{b}_i}
&=\mathbf{Tr}\Big[p^{\rho}p^{\sigma}M_{\rho\sigma}(\s{p}_1+m_{\tilde{G}})(B_i+Z_i\g_5)\s{p}_2(B_j-Z_j\g_5)\Big],\\
W_{\chi_i^+\chi_i^+}^G &=\mathbf{Tr}\Big[M_{\rho\sigma}(\s{p}_1+m_{\tilde{G}})\s{p}\g^{\rho}(T_i+Q_i\g_5)(\s{q}_3+m_{\chi})(S_i+P_i\g_5)\s{p}_2\\\nonumber
&\quad(S_j-P_j\g_5)(\s{q}_3+m_{\chi})(T_j-Q_j\g_5)\g^{\sigma}\s{p}\Big],
\end{align}
the functions $W_{\psi_a\psi_a}^G$  depend on the scalar products of the momenta $p, p_1, p_2, k, q_1, q_2$ and $q_3$, these functions will also be written as powers of the intermediate state masses, namely: 
\begin{equation}\label{eq:72:Wfunctionsgolstino}
W_{\psi_a\psi_a}^G=\text{w}_{1\psi_a\psi_a}^G+m_{\psi_a}\text{w}_{2\psi_a\psi_a}^G+m_{\psi_a}^2\text{w}_{3\psi_a\psi_a}^G.
\end{equation}
 All the full expressions for each function $\text{w}_{i\psi_a\psi_a}^G$ $\forall\,i=1,2,3$  can be foud in Appendix \ref{appendixB}. 
Again, the interferences terms for the goldstino are also written in the form:
\begin{equation}\label{eq:73:goldstino:interferences}
\mathcal{M}_{\psi_a}^{G\,\dagger}\mathcal{M}_{\psi_b}^G=\widetilde{C}_{\psi_a}\widetilde{C}_{\psi_b}P^*_{\psi_a}(q_a)P_{\psi_b}(q_b)W_{\psi_a\psi_b}^G.
\end{equation}
The functions $W_{\psi_a\psi_b}$ correspond to the traces involved in the interference terms, i.e. 
\begin{align}
W_{t\tilde{b}_i}^G &=\mathbf{Tr}\Big[M_{\rho\sigma}\s{p}_2P_R\g^{\sigma}(\s{q}_1+m_t)(A_{\tilde{t}}-B_{\tilde{t}}\g_5)(\s{p}_1+m_{\tilde{G}})(B_i+Z_i\g_5)p^{\rho}\Big], \\
W_{t \chi_i^+}^G &= \mathbf{Tr}\Big[M_{\rho\sigma}(\s{p}_1+m_{\tilde{G}})\s{p}\g^{\rho}(T_i+Q_i\g_5)(\s{q}_3+m_{\chi})(S_i+P_I\g_5)\\ \nonumber
 &\quad\s{p}_2P_R\g^{\sigma}(\s{q}_1+m_t)(A_{\tilde{t}}-B_{\tilde{t}}\g_5)\Big],\\
 W_{\chi_i^+\tilde{b}_i}^G &= \mathbf{Tr}\Big[M_{\rho\sigma}\s{p}_2(S_i-P_i\g_5)(\s{q}_3+m_{\chi})(T_i-Q_i\g_5)\g^{\rho}\s{p}(\s{p}_1+m_{\tilde{G}})(R_j+Z_j\g_5)p^{\sigma}\Big].
 \end{align}
The $W_{\psi_a\psi_b}^G$ functions also expressed as powers of the intermediate masses:
\begin{equation}
W_{\psi_a\psi_b}=\text{w}_{1\psi_a\psi_b}^G+m_{\psi_a}(\text{w}_{2\psi_a\psi_b}^G+m_{\psi_b}\text{w}_{3\psi_a\psi_b}^G)+m_{\psi_b}\text{w}_{4\psi_a\psi_b}^G.
\end{equation}
The full expressions for $\text{w}_{j\psi_a\psi_b}^G$ $\forall\,j=1,2,3,4$ can be found in the Appendix \ref{appendixB}.
\section{Numerical Results}\label{numericalsection}
The decay width is obtained by integrating the differential decay width over the dimensionless variables $x,\,y$  
which have limits given by 
$2\mu_G<x<1+\mu_{\tilde{G}}-\mu_W$ with $\mu_i=\frac{m_i^2}{m_{\tilde{t}_1}^2}$  and 
\begin{equation}\label{eq:206}
y_{\pm}=\frac{(2-x) \left(\mu _{\tilde{G}}+\mu _W-x+1\right)\pm\sqrt{x^2-4 \mu _{\tilde{G}}} \left(\mu _{\tilde{G}}-\mu _W-x+1\right)}{2 \left(\mu _{\tilde{G}}-x+1\right)},
\end{equation}
\begin{equation}\label{eq:205}
\Gamma=\int_{2\mu_G}^{1+\mu_G-\mu_W}\int_{y_-}^{y_+}\frac{m_{\tilde{t}_1}^2}{256\,\pi^3}\mid\overline{\mathcal{M}} \mid^2dy\,dx.
\end{equation}
After integrating numerically the expressions for the differential decay width, we obtain the values for the decay width, 
for a given set of parameters. We consider two values for the stop mass, $m_{\tilde{t}_1}=200\,GeV$ and $m_{\tilde{t}_1}=350\,GeV$, 
we also fix the chargino mass to be $m_{\chi_i^+}=200,\,500\,GeV$, while the sbottom mass is fixed to be 
$m_{\tilde{b}_i}=300,\,500\, GeV$. 

\begin{figure}[H]
\begin{center}
\includegraphics[scale=0.38]{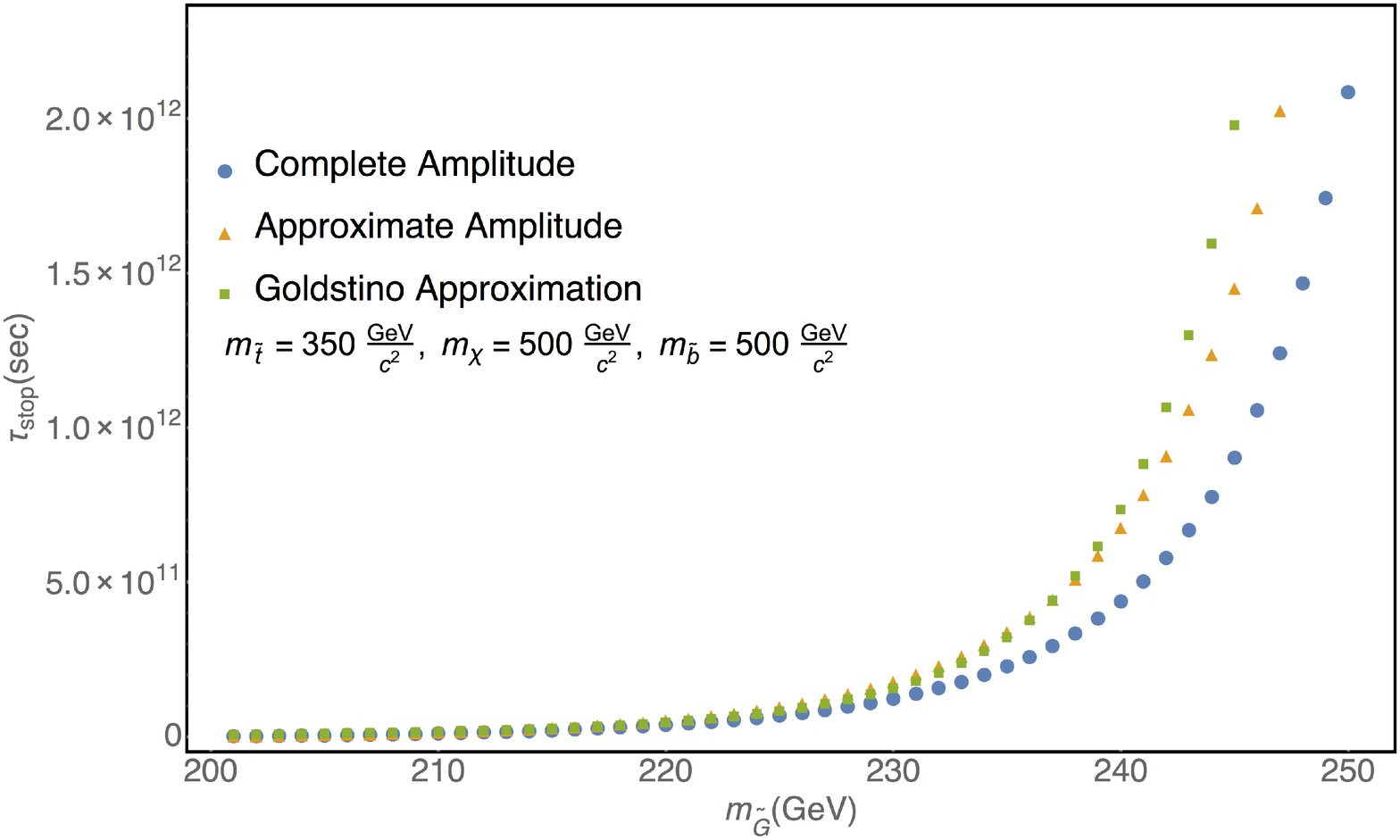}
\caption{Stop lifetime 1}
\label{fig:plot1}
\end{center}
\end{figure}

\begin{figure}[H]
\begin{center}
\includegraphics[scale=0.38]{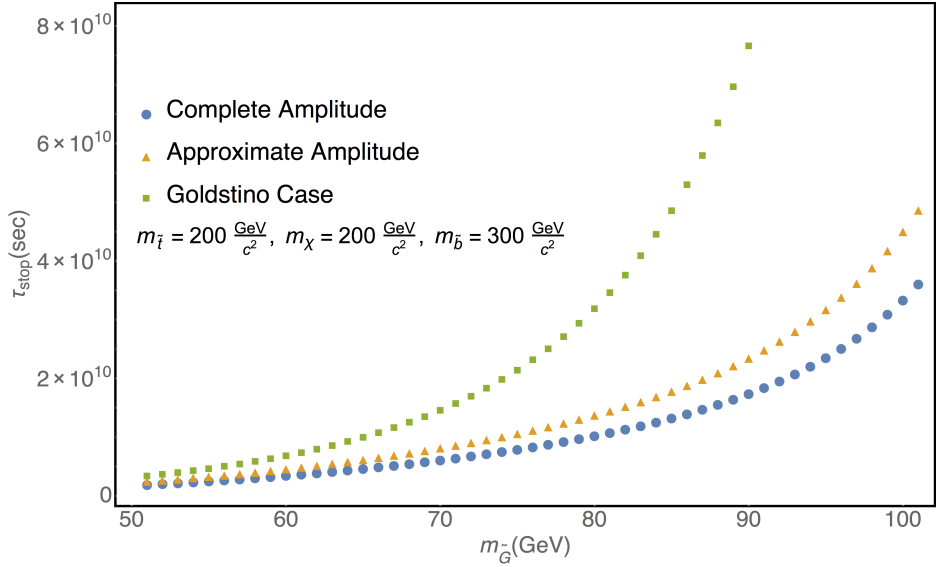}
\caption{Stop lifetime 2}
\label{fig:plot2}
\end{center}
\end{figure}

In Figures~[\ref{fig:plot1},\ref{fig:plot2}] we show the lifetime of the stop, as function of the gravitino mass, 
within the ranges 200-250 GeV for the case with $m_{\tilde{t}_1}=350\,GeV$,  and 50-100 GeV for $m_{\tilde{t}_1}=100\,GeV$.
We show the results for the case when one uses the full expression for  chargino-gravitino-W vertex  (circles), 
as well as the case when the partial inclusion of such vertex, as it was done in \cite{jlorenzo} (triangles) 
and in the limit of the goldstino approximation (squares). 
We noticed that for low gravitino masses ($m_{\widetilde{G}}\to  0$) the full gravitino result becomes almost 
indistinguishable from the goldstino case, while the partial gravitino result has also similar behavior. 
For larges gravitino masses ($m_{\widetilde{G}}\cong m_{\tilde{t}_1}$) the results for the stop lifetime using 
the full gravitinio and goldstino approximation could be very different, up to $O(50\%)$ different.

On the other hand, the values for the stop life-time using the full gravitino and partial gravitino limit
are very similar for low gravitino masses, while for the largest allowed masses the difference in results is at most
of order $O(50\%)$. 
The value of the lifetime obtained in all theses cases turns out to be of order $10^{7}\, -\,10^{12}$ sec, which 
results in an scenario with large stop lifetime that has very special signatures both at colliders and has also
important implications for cosmology, as it was discussed in ref. \cite{jlorenzo}.
For instance, regarding the effect on BBN, the  Stop $\tilde{t}_1$ have to form quasi stable sbaryons ($\tilde{t}_1 q q$) and mesinos ($\tilde{t}_1\bar{q}$), whose late decays could have affected the light element abundance obtained in BBN, while negatively charged stop sbaryons and mesinos could contribute to lower the Coulomb barrier for nuclear fusion process occurring in the BBN epoch. However, as argued in \cite{jlorenzo} the great majority of stop antisbaryons would have annihilated with ordinary baryons to make stop antimesinos and most stop mesinos and antimesinos would have annihilated. The only remnant would have been neutral mesinos which would be relatively innocuous, despite their long lifetime because they would not have important bound state effects. Further discussion of BBN issues of Ref. \cite{covi} divide the stop lifetime into regions that could have an effect, but the larges ones (which represent our results) do not pose problems for the success of BBN. Then, regarding the effect of late stop decay on the Cosmic Microwave Background (CMB), we have included some comments  in the text, to estimate the main effects. The arguments which read as follows:
Very long lifetimes ($\tau>10^{12}\,s$) would have been excluded if one uses the approximate results of Ref.~\cite{Feng2}, which present bounds on the lifetime $\tau$ (for the case when stau is the NLSP) using the constrain in the chemical potential $\mu<9\times10^{-5}$. However, it was discussed  in  Ref.~\cite{Durrer}, that a more precise calculation reduces the excluded region for lifetimes, ending  at about  $\tau \sim 10^{11}s-10^{12}s$. Thus, the region with very large stop lifetimes could also survive. Specific details that change from the stop decay  (3-body) as compared with stau  decays (2-body), such as the energy release or stop hadronization, will affect the calculation, but the numerical evaluation of 
such effect is beyond the scope of our paper.

\section{Conclusions}\label{conclusions}
In this paper we have calculated the  stop $\tilde{t}_1$ lifetime in MSSM scenarios  
where the massive gravitino is the lightest supersymmetric particle (LSP), 
and therefore is a viable dark matter candidate. The lightest stop $\tilde{t}_1$
corresponds to the next-to-lightest supersymmetric particle (NLSP). 
We have focused on the kinematical domain $m_{\widetilde{G}}+m_t>m_{\tilde{t}_1}>m_{\widetilde{G}}+m_b+m_W$, 
where the stop decay width is dominated by the mode $\tilde{t}_1\rightarrow b\,W\,\tilde{G}$.

The amplitiude for the full calculation of the stop 3-body decay width 
includes contributions from top, sbottom and chargino as intermediate states.
We have considered the full chargino-gravitino vertex, which improves
the calculation presented in ref. \cite{jlorenzo}.
Besides performing the full calculation with massive gravitino, we have also evaluated the 
stop decay lifetime for the limit when the gravitino can be approximated by the goldstino state. 
Our analytical results are conveniently expressed, in both cases, using an expansion in terms of 
the intermediate state mass, which helps in order  to identify the massless limit. 

We find that the results obtained with the full chargino vertex are not very different from the approximation used 
in ref. \cite{jlorenzo}, in fact they only differ approximately in a 50\%.
The comparison of the full numerical results with the ones 
obtained for the goldstino approximation, show that 
in the limit of low gravitino mass ($m_{\widetilde{G}}\ll m_{\tilde{t}_1}$) 
there is not a significant difference in values of the stop life-time obatined from each method. 
However, for $m_{\widetilde{G}} \lesssim m_{\tilde{t}_1}$ the difference 
in lifetime  could be as high as 50\%.
Numerical results for the stop lifetime give value of order $10^7\,-\,10^{12}$ sec, 
which makes the stop to behave like a quasi-stable state, 
which leaves special imprints for LHC search.
Our calculation shows that the inclusion of the neglected term somehow gives a decrease in the lifetime of the stop. However, it should be pointed out that the region of parameter space correspond to the NUHM model.
\appendix
\section{Analytical Expressions for Amplitudes with Gravitino in the final state}
\label{apendixA}
In this appendix we present  explicitly the full results for  the 10 $\text{w}_{\psi_a\psi_a}$ functions that arose from a convenient way to express the large traces that appear in the squared amplitudes [\ref{eq:14}], as well as the 18 $\text{w}_{\psi_a\psi_b}$ functions  in the interferences [\ref{eq:73:interferences}] of the 3-body stop $\tilde{t}_1$ decay with gravitino in the final state.  First, we shall present the contributions for the squared amplitudes, then we shall present the interferences.  
\subsection{Top Contribution}
For the averaged squared amplitude of the top quark  contribution,
the functions $\text{w}_{1tt},\, \text{w}_{2tt}$ and $\text{w}_{3tt}$ are:
\begin{align}\label{eq2:w1:top.old.amp}\nonumber
\text{w}_{1tt} & = \frac{4 a_1 h_1}{3 m_W^2
   m_{\tilde{G}}^2}
   \Bigl(f_2 (m_W^2
   (6 m_{\tilde{G}}^2+2
  h_5 m_{\tilde{t}_1}^2-q_1^{2})+6 f_3 m_W^2+4
   f_3^2)-2 f_1
   \bigl(m_{\tilde{G}}^2 \bigl(-\bigl(4 f_3\\ \nonumber
   &\quad+3
   m_W^2\bigr)\bigr)+f_2 (4 f_3+3
   m_W^2)+f_3 q_1^{2}\bigr)+(q_1^{2}-2
   m_{\tilde{G}}^2) (m_W^2
   m_{\tilde{G}}^2+3 f_3 m_W^2+2
   f_3^2)\\ 
   &\quad+4
   f_1^2
   (f_2-m_{\tilde{G}}^2)-2 m_W^2
   m_{\tilde{G}}^2h_5 m_{\tilde{t}_1}^2-4
   f_2^2 m_W^2\Bigr),
   \\ 
  \text{w}_{2tt} & = \frac{8 a_2 h_1}{3 m_W^2
   m_{\tilde{G}}}
(m_W^2
   (m_{\tilde{G}}^2+m_{\tilde{t}}^2-2
   f_2)-f_1 (4 f_3+3 m_W^2)+3 f_3
   m_W^2+2 f_1^2+2
   f_3^2)\label{eq2:w2top.old.amp},
\\
\text{w}_{3tt} & = \frac{4 a_3 h_1}{3 m_W^2
   m_{\tilde{G}}^2}  
   (m_W^2
   (f_2-m_{\tilde{G}}^2)-3 f_3 m_W^2-2
   f_3^2+2 f_1 f_3)\label{eq3:w3top.old.amp}.
\end{align}
The functions $f_1,\,f_2$ and $f_3$ are functions of the variables $x$ and $y$  that were defined previously in Section \ref{numericalsection}, they are $f_1=\frac{m_{\tilde{t}_1}^2}{2}y,\,f_2=\frac{m_{\tilde{t}_1}^2}{2}x,\,f_3=\frac{m_{\tilde{t}_1}^2}{2}(-1-\mu_{\widetilde{G}}-\mu_{W}+x+y)$,
with $\mu_{\widetilde{G}}=\frac{m_{\widetilde{G}}^2}{m_{\tilde{t}_1}^2}$ and $\mu_{W}=\frac{m_{W}^2}{m_{\tilde{t}_1}^2}$. 
We  have also used in [\ref{eq2:w1:top.old.amp}-\ref{eq3:w3top.old.amp}] the following substitutions $h_1=(f_2^2-m_{\tilde{G}}^2m_{\tilde{t}_1}^2)$, $a_1=(A_{\tilde{t}}-B_{\tilde{t}})^2$, $a_2=A_{\tilde{t}}^2-B_{\tilde{t}}^2$ and $a_3=(A_{\tilde{t}}+B_{\tilde{t}})^2$, 
with $A_{\tilde{t}}=\cos\theta_{\tilde{t}}+\sin \theta_{\tilde{t}}$ and $B_{\tilde{t}}=\cos\theta_{\tilde{t}}-\sin \theta_{\tilde{t}}$.
\subsection{Sbottom Contribution}
For the averaged squared amplitude of the  squark sbottom contribution,  the function $\text{w}_{1\tilde{b}_i\tilde{b}_i}$ is:
\begin{equation}\label{eq4:w4:sbotom.old}
 \text{w}_{1\tilde{b}_i\tilde{b}_i} = \frac{8 D_{ij1}
   h_2 h_3
   ((f_2-f_3){}^2-q_2^{2}
   m_{\tilde{G}}^2)
   }{3 m_W^2 m_{\tilde{G}}^2}.
\end{equation}
With $h_2=f_2-f_3-m_{\tilde{G}}^2$ and $h_3=f_1^2-m_W^2m_{\tilde{t}_1}^2$.
We have done in the amplitude [\ref{eq:55:sbotom:oldamplitude}] the following substitution $a_iP_R+b_iP_L=\frac{1}{2}(R_i+Z_i\g_5)$ such that $D_{ij1}=R_iR_j+Z_iZ_j,\text{with}\,R_i=a_i+b_i,\,Z_i=a_i-b_i,\,R_j=a_j+b_j$ and $Z_j=a_j-b_j$, and 
 with $a_i=(\sin\theta_{\tilde{b}},\cos\theta_{\tilde{b}}),\,b_i=(\cos\theta_{\tilde{b}},-\sin\theta_{\tilde{b}})$ and $\kappa_i=(\cos\theta_{\tilde{t}}\cos\theta_{\tilde{b}},-\cos\theta_{\tilde{t}}\cos\theta_{\tilde{b}})$.

\subsection{Partial Chargino Contribution ($\mathcal{M}_{\chi_i^+}^0$)}
For the averaged squared amplitude of the chargino contribution, the functions $\text{w}_{k\chi_i^+\chi_i^+}^0,\,\forall\,k=1,2,3$ are as follows
\begin{align}\label{eq6:w5:old.amp}\nonumber
\text{w}_{1\chi_{i}^{+}\chi_{i}^{+}}^0 &=-\frac{8 \Sigma_{\text{ij1}}h_4}{3 m_W^2
   m_{\tilde{G}}^2} 
   \Bigl((m_{\tilde{G}}^2+f_3) (2
   (m_{\tilde{G}}^2+m_W^2)+4
   f_3-q_3^{2})\\ 
   &\quad+f_2 (-2 m_{\tilde{G}}^2-2
   f_3+q_3^{2})-2 f_1
   (m_{\tilde{G}}^2+f_3)\Bigr),
 \\ 
   \text{w}_{2\chi_{i}^{+}\chi_{i}^{+}}^0 &=-\frac{8 h_4(\Sigma_{\text{ij1}}+\Sigma_{\text{ij2}})(h_5-f_1-f_2)}{3
   m_W^2 m_{\tilde{G}}} 
   \label{eq6:w6:old.amp},
   \\ 
   \text{w}_{3\chi_{i}^{+}\chi_{i}^{+}}^0 &= \frac{8 \Sigma _{\text{ij3}}h_4h_2}{3
   m_W^2 m_{\tilde{G}}^2}
  \label{eq6:w8:old.amp},
   \end{align}
   with $h_4=2 m_W^2
   m_{\tilde{G}}^2+f_3^2$ and $h_5=m_{\tilde{G}}^2+2f_3+m_W^2$, we  have also used the following substitutions $\Sigma_{ij1}=(S_iS_j+P_iP_j)(V_iV_j-\Lambda_i\Lambda_j)-(S_iP_j+P_iS_j)(\Lambda_iV_j-V_i\Lambda_j)$, $\Sigma_{ij2}=(S_iS_j+P_iP_j)(V_iV_j-\Lambda_i\Lambda_j)+(S_iP_j+P_iS_j)(\Lambda_iV_j-V_i\Lambda_j)$, $\Sigma_{ij3}=(S_iS_j+P_iP_j)(V_iV_j+\Lambda_i\Lambda_j)+(S_iP_j+P_iS_j)(\Lambda_iV_j+V_i\Lambda_j)
$, 
   with $V_{i}=V_{i2}\sin{\beta}+U_{i2}\cos\beta$ and $\Lambda_{i}=V_{i2}\sin{\beta}-U_{i2}\cos\beta$. 
For the low-to-moderate range of $\tan\beta$ we have:
\begin{align}
S_1&=\frac{1}{2}\left(-g_2\cos\phi_L+\frac{g_2m_t\sin\phi_L\sin\theta_{\tilde{t}}}{\sqrt{2}m_W\sin\beta}\right)\label{eq:68},\\
P_1&=\frac{1}{2}\left(-g_2\cos\phi_L-\frac{g_2m_t\sin\phi_L\sin\theta_{\tilde{t}}}{\sqrt{2}m_W\sin\beta}\right)\label{eq:69},
\end{align}
where $\cos\phi_L,\pm\sin\phi_L$ are elements of the matrix $V$ that diagonalizes the chargino mass matrix, expressions for $S_2$ and $P_2$ may be obtained by replacing $\cos\phi_L\rightarrow -\sin\phi_L$ and $\sin\phi_L\rightarrow\cos\phi_L$ in [\ref{eq:68}] and [\ref{eq:69}].
\subsection{Full Chargino Contribution ($\widetilde{\mathcal{M}}_{\chi_i^+}$)}
For the averaged squared amplitude $\widetilde{\mathcal{M}}_{\chi_i^+}$ of the chargino contribution,  the  functions $\text{w}_{k\chi_i^+\chi_i^+}$ $\forall\,k=1,2,3$, are:
\begin{align}
 \text{w}_{1\chi_i^+\chi_i^+} & = q_3^2 P_{\text{ij1}}h_7\label{eq:19:w71.chargino-chargino.new},
   \\ 
   \text{w}_{2\chi_i^+\chi_i^+} & = \frac{16  m_{\tilde{G}}(P_{\text{ij1}}+P_{\text{ij2}})}{3 m_W^2}
   (h_5-f_1-f_2)
   (2 f_1^2-5  m_W^2
  m_{\tilde{t}_1}^2),\label{eq:20:w18:chargino-chargino.new}
  \\ 
 \text{w}_{3\chi_i^+\chi_i^+} & = P_{\text{ij2}}h_7,
   \label{eq:22:w20:chargino-chargino.new}
\end{align}
where we have defined
\begin{align}
h_7 &=\frac{16}{3 m_W^2 m_{\tilde{G}}^2}\left(2 f_1 (f_2 (2
   (f_2-f_3) f_3-m_{\tilde{G}}^2 (2
   f_3+m_W^2))-f_3 m_{\tilde{G}}^2
   m_{\tilde{t}_1}^2)\nonumber \right.\\ 
   &\quad\left.+h_2 (2 f_2^2 m_W^2-m_{\tilde{t}_1}^2 h_6)+f_1^2 (4 f_2 m_{\tilde{G}}^2-2
   m_{\tilde{G}}^4)\right).
\end{align}
With $h_6=3 m_W^2
   m_{\tilde{G}}^2+2
   f_3^2$, we have used the  substitution $V_{i1}P_R-U_{i1}P_L=T_i+Q_i\g_5$ in the first term of the interaction vertex $V_6(\chi_i^+\, W\,\tilde{G})$,
we have also done the following substitutions in the functions [\ref{eq:19:w71.chargino-chargino.new}-\ref{eq:22:w20:chargino-chargino.new}]:
\begin{align}
P_{ij1} &= (S_iS_j+P_iP_j)(T_iT_j+Q_iQ_j)-(S_iP_j+P_iS_j)(T_iQ_j+Q_iT_j),\\
P_{ij2} &= (S_iS_j+P_iP_j)(T_iT_j+Q_iQ_j)+(S_iP_j+P_iS_j)(T_iQ_j+Q_iT_j).
\end{align}
\subsection{Interference Terms}
\subsubsection*{$\mathcal{M}_{\chi_i^+}^{0\dagger}
\mathcal{\widetilde{M}}_{\chi_i^+}$ Interference}
For the interference term $\mathcal{M}_{\chi_i^+}^{0\dagger}\mathcal{\widetilde{M}}_{\chi_i^+}$,
the  $\widetilde{\text{w}}_{k\chi_i^+\chi_i^+}$ functions $\forall\,k=1,2,3,4$, are:
\begin{align}\nonumber
\widetilde{\text{w}}_{1{\chi_{i^+}\chi_{i^+}}} &= \frac{16 S_{\text{ij1}}}{3 m_W^2 m_{\tilde{G}}}
 \Bigl(f_1^2
 (m_{\tilde{G}}^2 (8 f_3+2m_W^2-q_3^{2})+4f_3^2)
   +f_1 \bigl(4 f_2 f_3 (2m_{\tilde{G}}^2+f_3-q_3^{2}) \\ \nonumber
   &\quad-(m_{\tilde{G}}^2 (4 f_3+m_W^2)+2
   f_3^2) (2(m_{\tilde{G}}^2+m_W^2)+4
   f_3-q_3^{2})\bigr) 
   \\ 
   &\quad+f_2 m_W^2
   (-m_{\tilde{G}}^2 (2 f_2-4 f_3-2
   m_W^2+q_3^2)+2 m_{\tilde{G}}^4+f_2
   q_3^2)+f_3^2 q_3^2
  m_{\tilde{t}_1}^2\Bigr),\label{eq:24:w21:chargino-chargino.oldnew}
 \\ 
 \nonumber\widetilde{\text{w}}_{2{\chi_{i^+}\chi_{i^+}}}&=\frac{16 (S_{\text{ij2}}+S_{\text{ij3}})}{3 m_W^2 m_{\tilde{G}}^2} \Bigl(m_{\tilde{t}}^2
   (-f_3 m_{\tilde{G}}^2 (f_3-3
   m_W^2)+2 m_W^2 m_{\tilde{G}}^4-2
   f_3^3)\\ \nonumber
   &\quad+2 f_1 \bigl(f_3
   m_{\tilde{G}}^2 h_5+2 f_2 (m_W^2
   m_{\tilde{G}}^2+f_3^2)\bigl)-
   f_2 m_W^2 (5 m_{\tilde{G}}^2
  h_5+f_2
   (2 f_3-3
   m_{\tilde{G}}^2)) \\ 
      &\quad-f_1^2
   (4 f_3
   m_{\tilde{G}}^2+m_{\tilde{G}}^4)\Bigr), 
   \label{eq:25:w22:chargino-chargino.oldnew}
 \\ 
 \widetilde{\text{w}}_{3{\chi_{i^+}\chi_{i^+}}} &= -\frac{16 S_{\text{ij4}}}{3 m_W^2 m_{\tilde{G}}} 
 \Bigl(f_2 m_W^2
   (f_2-m_{\tilde{G}}^2)+f_1
   (m_{\tilde{G}}^2 (4 f_3+m_W^2)+2
   f_3 (f_3-2
   f_2)) \nonumber \\ 
   &\quad-f_1^2
   m_{\tilde{G}}^2+f_3^2
  m_{\tilde{t}_1}^2\Bigr),
  \label{eq:27:w24:chargino-chargino.oldnew}
\end{align}
In order to have control in the calculations with huge expressions,  we have done the following substitutions in the functions [\ref{eq:24:w21:chargino-chargino.oldnew}-\ref{eq:27:w24:chargino-chargino.oldnew}]:
\begin{align}\label{eq:106}
S_{ij1} & =  (S_iS_j+P_iP_j)(T_iV_j+Q_i\Lambda_j)-(S_iP_j+P_iS_j)(Q_iV_j+T_i\Lambda_j),\\
S_{ij2} & =  (S_iS_j+P_iP_j)(T_i\Lambda_j+Q_iV_j)-(S_iP_j+P_iS_j)(Q_i\Lambda_j+T_iV_j),\\
S_{ij3} & =  (S_iS_j+P_iP_j)(T_i\Lambda_j+Q_iV_j)+(S_iP_j+P_iS_j)(Q_iV_j+T_i\Lambda_j),\\
S_{ij4} & =  (S_iS_j+P_iP_j)(T_iV_j+Q_i\Lambda_j)+(S_iP_j+P_iS_j)(Q_iV_j+T_i\Lambda_j).
\end{align}
\subsubsection*{$\mathcal{M}_{\chi_i^+}^{0\dagger}\mathcal{M}_{\tilde{b}_i}$ Interference}
For the interference term $\mathcal{M}_{\chi_i^+}^{0\dagger}\mathcal{M}_{\tilde{b}_i}$, 
the functions  $\text{w}_{j\chi_i^+\tilde{b}_i}$ $\forall\,j=1,2$ are:
\begin{align}\label{eq:16:w15:chargino-sbotom}\nonumber
\text{w}_{1\chi_i^+\tilde{b}_i}&=-\frac{4 \eta _{\text{ij1}}}{3 m_W^2 m_{\tilde{G}}} \Bigl(-f_1
   \bigl(m_{\tilde{t}}^2 (m_W^2
   m_{\tilde{G}}^2+f_3^2)-2 f_3
   f_2 (m_{\tilde{G}}^2+3 f_3-m_W^2)\\ \nonumber
   &\quad+2
   f_3^2 h_5+f_2^2 (2
   f_3-m_W^2)\bigr)+m_W^2 \bigl(m_{\tilde{t}}^2
   ºbigl(m_{\tilde{G}}^2 (-2 f_2+4
   f_3+m_W^2)\\ \nonumber
   &\quad+2
   m_{\tilde{G}}^4+f_3^2\bigr)+f_2
   (-f_2 (2 m_{\tilde{G}}^2+6
   f_3+m_W^2)+2 f_3 h_5+2
   f_2^2)\bigr)\\ 
   &\quad+f_1^2 (-m_{\tilde{G}}^2 (-2 f_2+4
   f_3+m_W^2)-2 m_{\tilde{G}}^4+2
   f_3^2)+f_1^3
   m_{\tilde{G}}^2\Bigr),
   \\ 
  \text{w}_{2\chi_i^+\tilde{b}_i} &=\frac{8 \eta _{\text{ij2}}h_2}{3 m_W^2 m_{\tilde{G}}^2}
(m_W^2
   (m_{\tilde{G}}^2h_5 m_{\tilde{t}_1}^2+f_2
   (f_3-f_2))-f_1^2
   m_{\tilde{G}}^2+f_1 (f_2-f_3)
   f_3).
   \label{eq:17:w16:chargino-sbotom}
   \end{align}
In the functions [\ref{eq:16:w15:chargino-sbotom},\ref{eq:17:w16:chargino-sbotom}], we have done the following substitutions:
 \begin{align}\label{eq:108}
 \eta_{ij1} &= R_j(\Lambda_iS_i-V_iP_i)+Z_j(\Lambda_iP_i-V_iS_i),\\
 \eta_{ij2} &= R_j(\Lambda_iS_i+V_iP_i)+Z_j(\Lambda_iP_i+V_iS_i).\\
 \end{align}

\subsubsection*{$\mathcal{M}_t^{\dagger}\mathcal{M}_{\chi_i^+}^0$ Interference}
For the interference term $\mathcal{M}_t^{\dagger}\mathcal{M}_{\chi_i^+}^0$ 
the functions $\text{w}_{jt\chi_i^+}$ $\forall\,j=1,2,3,4$ are:
\begin{align}\nonumber
\text{w}_{1t\chi_i^+} &=\frac{2 \Omega _{\text{i1}}}{3
   m_W^2 m_{\tilde{G}}}
 \Bigl(2 f_1
   \bigl(m_{\tilde{t}}^2 (m_{\tilde{G}}^2
   (f_3+2
   m_W^2)-f_3^2)-f_2
   (m_{\tilde{G}}^2h_8+2
   f_3 (3 f_3+m_W^2))\\ \nonumber
   &\quad-f_3
   m_{\tilde{G}}^2 h_5+f_2^2 (2 f_3-3
   m_W^2)\bigr)+m_{\tilde{t}}^2 \bigl(-m_W^2
   m_{\tilde{G}}^2 (-4 f_2+6 f_3+m_W^2)\\ \nonumber &\quad-4
   m_W^2 m_{\tilde{G}}^4+f_3 ((2
   f_2+f_3) m_W^2+2 f_3
   (f_3-f_2))\bigr)\\ \nonumber
   &\quad+f_1^2 (m_{\tilde{G}}^2 (-4 f_2+10 f_3+3
   m_W^2)+4 m_{\tilde{G}}^4+8 f_2 f_3)+f_2
   \bigl(2 f_3^2
   (m_{\tilde{G}}^2+m_W^2)\\ 
   &\quad+f_2 m_W^2
   (4 m_{\tilde{G}}^2-4 f_2+m_W^2)+4 f_2
   f_3 m_W^2+4 f_3^3\bigr)-6
   f_1^3 m_{\tilde{G}}^2\Bigr),
   \label{eq11:w11:topchargino.int}\\
   \text{w}_{2t\chi_i^+} &= \frac{4 \Omega _{\text{i2}}}{3 m_W^2
   m_{\tilde{G}}^2}
 \Bigl(-m_{\tilde{t}}^2
   (f_2 (f_3^2-2 m_W^2
   m_{\tilde{G}}^2)+2 m_W^2
   m_{\tilde{G}}^4)
   +f_1 \bigl(f_3
   m_{\tilde{G}}^2 m_{\tilde{t}_1}^2\nonumber\\ \nonumber
   &\quad+m_{\tilde{G}}^4
   (m_W^2-f_3)-f_2 m_{\tilde{G}}^2 (2
   f_3+m_W^2)+2 f_2^2
   f_3\bigr)\nonumber \\ 
   &\quad
   +f_2 (m_{\tilde{G}}^2 ((2
   f_2-f_3)
   m_W^2+f_3^2)+f_2 (f_3-2
   f_2) m_W^2)
   +2 f_1^2
   m_{\tilde{G}}^2
   (m_{\tilde{G}}^2-f_2)\Bigr),  
   \label{eq12:w12:topchargino.int}
   \\ 
   \text{w}_{3t\chi_i^+} &= \frac{2 \Omega _{\text{i3}}}{3 m_W^2
   m_{\tilde{G}}} \Bigl(m_{\tilde{t}}^2
   h_9-f_1
   m_{\tilde{G}}^2 (f_1-2 f_3+2 m_W^2)-3
   f_2^2 m_W^2+2 f_2 f_3
   \left(m_W^2-f_3\right)\Bigr),
   \label{eq13:w13:topchargino.int}
   \\ 
   \text{w}_{4t\chi_i^+} &= -\frac{2 \Omega _{\text{i4}}}{3
   m_W^2 m_{\tilde{G}}^2}
 \Bigl(-m_{\tilde{G}}^2
  m_{\tilde{t}_1}^2 (4 f_3
   m_W^2+h_9)-2 f_1 \bigl(f_3
   m_{\tilde{G}}^2 h_5 
    +f_2 (2
   f_3^2-m_W^2
   m_{\tilde{G}}^2)\bigr) \nonumber \\
  &\quad  +2 f_2f_3^2 h_5+f_2^2 m_W^2(3
   m_{\tilde{G}}^2+2 f_3)+f_1^2
   (4 f_3
   m_{\tilde{G}}^2+m_{\tilde{G}}^4)\Bigr).
   \label{eq14:w14:topchargino.int}
   \end{align}
With $h_8=2f_3-m_W^2$ and $h_9=3 m_W^2
   m_{\tilde{G}}^2+f_3^2$. We have done the following substitutions  in the functions [\ref{eq11:w11:topchargino.int}-\ref{eq14:w14:topchargino.int}]:
    $\Omega_{i1}=(A_{\tilde{t}}-B_{\tilde{t}})(S_i-P_i)(\Lambda_i+V_i)$, $\Omega_{i2}=(A_{\tilde{t}}-B_{\tilde{t}})(S_i-P_i)(\Lambda_i-V_i)$, $\Omega_{i3} = (A_{\tilde{t}}+B_{\tilde{t}})(S_i-P_i)(\Lambda_i-V_i)$ and $\Omega_{i4}=(A_{\tilde{t}}+B_{\tilde{t}})(S_i-P_i)(\Lambda_i+V_i)$.
\subsubsection*{$\mathcal{M}_t^{\dagger}\mathcal{M}_{\tilde{b}_i}$ Interference}
For the interference term $\mathcal{M}_t^{\dagger}\mathcal{M}_{\tilde{b}_i}$, 
the functions $\text{w}_{jt\tilde{b}_i}$ $\forall\,j=1,2$ are:
\begin{align}\nonumber
\text{w}_{1t\tilde{b}_i} &= \frac{2 \left(\Delta _{\text{i1}}+\Delta_{\text{i2}}\right)}{3 m_W^2 m_{\tilde{G}}^2}
 \Bigl(f_1^2
   \bigl(2 f_2 m_{\tilde{G}}^2 (-2
   m_{\tilde{t}_1}^2+h_8)-2
   m_{\tilde{G}}^2 m_{\tilde{t}_1}^2 (f_3-2
   m_{\tilde{G}}^2)+m_{\tilde{G}}^4 h_8  \\ \nonumber
   &\quad -4 f_2^2
   (m_{\tilde{G}}^2+f_3)+4
   f_2^3\bigr)+2 f_1
   \bigl(m_{\tilde{t}_1}^2 \bigl(-m_{\tilde{G}}^4
   (f_3-2 m_W^2)+f_3 m_W^2
   m_{\tilde{G}}^2\\ \nonumber
   &\quad+ f_2 f_3 (f_3-f_2)\bigr)+f_3 m_{\tilde{G}}^2
  m_{\tilde{t}_1}^4+f_2 \bigl(f_3 m_{\tilde{G}}^2
   (f_2-f_3+m_W^2)-m_W^2
   m_{\tilde{G}}^4  \\ \nonumber
   &\quad +f_2 (f_3-f_2)
   m_W^2 \bigr)\bigr)+m_W^2 \bigl(m_{\tilde{t}_1}^2
   \bigl(m_{\tilde{G}}^4 (2
   f_2+m_W^2)+(4 f_2^2-4 f_3
   f_2-f_3^2) m_{\tilde{G}}^2 \\ 
   &\quad -2
   f_2 (f_2-f_3)^2\bigr)-2
   m_{\tilde{G}}^2 m_{\tilde{t}_1}^4 (2
   m_{\tilde{G}}^2-f_2+f_3)-f_2^2
   m_{\tilde{G}}^2 (2 f_2-h_8)\bigr) \nonumber
\\ 
&\quad+4 f_1^3
   m_{\tilde{G}}^2
   (f_2-m_{\tilde{G}}^2)\Bigr),
   \label{eq8:w9:topsbotom.int.}
   \\ 
   \text{w}_{2t\tilde{b}_i} &= \frac{2 \left(\Delta_{\text{i1}}-\Delta_{\text{i2}}\right)}{3 m_W^2 m_{\tilde{G}}} 
   \Bigl(f_1^2
   (2 f_2^2-m_{\tilde{G}}^2 (2 m_{\tilde{t}_1}^2+h_8))+f_1
   \bigl(m_{\tilde{t}}^2 (m_{\tilde{G}}^2
   h_8-f_3^2)\nonumber \\ \nonumber
   &\quad +2 f_2
   (m_W^2 m_{\tilde{G}}^2-f_3
   m_W^2+f_3^2)-f_2^2 (2
   f_3+m_W^2)\bigr)
   +m_W^2
   \bigl(m_{\tilde{t}}^2 (-m_{\tilde{G}}^2
   \bigl(2 f_2\\ 
   &\quad +m_W^2)-2 f_2^2+2
   f_3 f_2+f_3^2)  +2
   m_{\tilde{G}}^2
   m_{\tilde{t}_1}^4+f_2^2 (2 f_2-h_8)\bigr)+f_1^3
   m_{\tilde{G}}^2\Bigr).
   \label{eq9:w10:topsbotom.int.}
   \end{align}
with $\Delta_{i1}=(R_i-Z_i)A_{\tilde{t}}$ and $\Delta_{i2}=(Z_i-R_i)B_{\tilde{t}}$.
\subsubsection*{$\widetilde{\mathcal{M}}_{\chi_i^+}^{\dagger}\mathcal{M}_{\tilde{b}_i}$ Interference}
For the interference term $\widetilde{\mathcal{M}}_{\chi_i^+}^{\dagger}\mathcal{M}_{\tilde{b}_i}$,
the functions $\widetilde{\text{w}}_{j\chi_i^+\tilde{b}_i}$ $\forall\,j=1,2$  are:
\begin{align}
\widetilde{\text{w}}_{1\chi_i^+\tilde{b}_i}&=\frac{8 f_1 C_{\text{ij1}}}{3 m_W^2
   m_{\tilde{G}}^2}
 \Bigl(2 f_1 \bigl(f_3
   m_{\tilde{G}}^2
   (m_{\tilde{t}}^2+h_5)+f_2 (2 m_W^2
   m_{\tilde{G}}^2+m_{\tilde{G}}^4-2
   f_3^2)\nonumber \\ \nonumber
   &\quad +f_2^2
   (2
   f_3-m_{\tilde{G}}^2)\bigr)-m_{\tilde{t}}^2
   \bigl(m_{\tilde{G}}^4 (2
   f_3+m_W^2)+m_{\tilde{G}}^2 \bigl(f_3 (3
   f_3+4 m_W^2) \\ \nonumber
   &\quad-2 f_2
   (f_3+m_W^2)\bigr)+2
   (f_2-f_3)
   f_3^2\bigr)-f_2 m_W^2
   \bigl(m_{\tilde{G}}^2 (-3 f_2+4 f_3+2
   m_W^2) \\
   &\quad+2 m_{\tilde{G}}^4+2 f_2
   (f_2-f_3)\bigr)+f_1^2
   (m_{\tilde{G}}^4-4 f_2
   m_{\tilde{G}}^2)\Bigr),\label{eq:29:w25:charginonew-sbotom} \\ 
   \widetilde{\text{w}}_{2\chi_i^+\tilde{b}_i}&=-\frac{16 f_1 C_{\text{ij2}}
   h_2 (-f_3
  h_5 m_{\tilde{t}_1}^2-f_2 m_W^2+f_1
   (f_2+f_3))}{3 m_W^2 m_{\tilde{G}}}.\label{eq:30:w26:charginonew-sbotom} 
   \end{align}
 We have done the following substitutions in the functions [\ref{eq:29:w25:charginonew-sbotom},\ref{eq:30:w26:charginonew-sbotom}],
\begin{align}\label{eq:109}\nonumber
C_{ij1} &= T_i(R_jS_i+Z_j P_i)-Q_i(R_jP_i+Z_jS_i),\\
C_{ij2} &= T_i(R_jS_i+Z_j P_i)+Q_i(R_jP_i+Z_jS_i).
\end{align}
\subsubsection*{$\mathcal{M}_t^{\dagger}\mathcal{\widetilde{M}}_{\chi_i^+}$ Interference}
For the interference term $\mathcal{M}_t^{\dagger}\mathcal{\widetilde{M}}_{\chi_i^+}$, 
the functions $\widetilde{\text{w}}_{jt\chi_{i}^+}$ $\forall\,j=1,2,3,4$  are:
\begin{align}\nonumber
\widetilde{\text{w}}_{1t\chi_{i}^+} &= \frac{8 R_{\text{i1}}}{3 m_W^2
   m_{\tilde{G}}^2}
 \Bigl(-m_{\tilde{G}}^2
   m_{\tilde{t}_1}^4 (4 f_3
   m_W^2+h_9)
   +m_{\tilde{t}}^2
   \bigl(f_2^2 m_W^2 (3
   m_{\tilde{G}}^2+4 f_3) +m_{\tilde{G}}^2
   \bigl(m_W^2 m_{\tilde{G}}^2\\ \nonumber
   &\quad-f_3^2\bigr)h_5+2 f_2 h_5 (m_W^2
   m_{\tilde{G}}^2+f_3^2)\bigr)  +
   f_1^2 \bigl(m_{\tilde{G}}^2
   m_{\tilde{t}_1}^2 (m_{\tilde{G}}^2+4
   f_3)-3 m_{\tilde{G}}^4
  h_5\\ \nonumber
   &\quad+f_2
   (4 m_W^2 m_{\tilde{G}}^2+6
   m_{\tilde{G}}^4)
    +f_2^2 (8
   f_3-4 m_{\tilde{G}}^2)\bigr)+2 f_1
   \bigl(m_{\tilde{t}}^2 \bigl(-(2
   m_{\tilde{G}}^4 (f_3+m_W^2)\\ \nonumber
   &\quad +(f_3-2
   f_2) m_{\tilde{G}}^2 (f_3+m_W^2) +2
   f_2 f_3^2)\bigr)
   -2 f_2
   (f_2-m_{\tilde{G}}^2) (f_3
  h_5+f_2
   m_W^2)\bigr)\\ 
   &\quad-f_2^2 m_W^2
   (m_{\tilde{G}}^2+2 f_2)
   h_5 +f_1^3 (6
   m_{\tilde{G}}^4-8 f_2
   m_{\tilde{G}}^2)\Bigr),
   \label{eq:32:w27:charginonew-top}
   \\ 
   \widetilde{\text{w}}_{2t\chi_{i}^+} &= \frac{8 R_{\text{i2}}}{3
   m_W^2 m_{\tilde{G}}}
 \Bigl(2 f_1
   (m_{\tilde{t}}^2 (m_{\tilde{G}}^2
   (f_3+2
   m_W^2)-f_3^2)-2 f_2
   (f_3 h_5+f_2
   m_W^2)) \nonumber\\ \nonumber
   &\quad +h_5 (m_{\tilde{t}}^2
   (f_3^2-m_W^2
   m_{\tilde{G}}^2)+f_2^2
   m_W^2)+f_1^2
   \bigl(m_{\tilde{G}}^2 (-2 f_2+6 f_3+3
   m_W^2) \\ 
   &\quad+3 m_{\tilde{G}}^4+8 f_2 f_3\bigr)-6
   f_1^3 m_{\tilde{G}}^2\Bigr), 
   \label{eq:33:w28:charginonew-top}\\
   \widetilde{\text{w}}_{3t\chi_{i}^+} &= \frac{8 R_{\text{i3}}}{3
   m_W^2 m_{\tilde{G}}^2}
 \Bigl(2 f_1 (-f_3
   m_{\tilde{G}}^2h_5 m_{\tilde{t}_1}^2+2 f_2
   m_{\tilde{G}}^2 (f_3-m_W^2)-2 f_3
   f_2^2)\nonumber  \\ \nonumber
   &\quad+m_{\tilde{t}}^2 (2
   f_2 (m_W^2
   m_{\tilde{G}}^2+f_3^2)-f
   _3^2 m_{\tilde{G}}^2+m_W^2
   m_{\tilde{G}}^4)-f_2^2 m_W^2
   (m_{\tilde{G}}^2+2 f_2-4
   f_3) \\ 
   &\quad+f_1^2 (4 f_2
   m_{\tilde{G}}^2-3 m_{\tilde{G}}^4)\Bigr), \label{eq:34:w29:charginonew-top}
\\ 
\widetilde{\text{w}}_{4t\chi_{i}^+} &= \frac{8 R_{\text{i4}}}{3 m_W^2 m_{\tilde{G}}}
 \Bigl(m_{\tilde{t}}^4h_9-m_{\tilde{t}}^2 \bigl(-m_{\tilde{G}}^2
   (f_3^2-2 f_2 m_W^2)+m_W^2
   m_{\tilde{G}}^4+f_2 ((3 f_2-4 f_3)
   m_W^2\nonumber  \\ \nonumber
   &\quad +2 f_3^2)\bigr)+2 f_1
   \bigl(f_2 (2 m_{\tilde{G}}^2
   (m_W^2-f_3)+f_2 (2
   f_3-m_W^2)) -m_{\tilde{t}}^2
   (m_{\tilde{G}}^2 (m_W^2-2 f_3)\\ 
   &\quad +2
   f_2 f_3)\bigr)+f_1^2 (-6
   f_2 m_{\tilde{G}}^2-m_{\tilde{G}}^2
  m_{\tilde{t}_1}^2+3 m_{\tilde{G}}^4+4
   f_2^2) +f_2^2
   m_W^2 (m_{\tilde{G}}^2+2 f_2-4
   f_3)\Bigr).
   \label{eq:34:w30:charginonew-top}
   \end{align}
 We have done the following substitutions in the functions [\ref{eq:32:w27:charginonew-top}
-\ref{eq:34:w30:charginonew-top}]
 $R_{\text{i1}}=(A_{\tilde{t}}-B_{\tilde{t}})(S_i+P_i)(T_i-Q_i)$, $ R_{\text{i2}}=(A_{\tilde{t}}+B_{\tilde{t}})(S_i-P_i)(T_i+Q_i)$, $ R_{\text{i3}}=(A_{\tilde{t}}-B_{\tilde{t}})(S_i+P_i)(T_i+Q_i)$ and $ R_{\text{i4}}= (A_{\tilde{t}}+B_{\tilde{t}})(S_i-P_i)(T_i-Q_i)$.
\section{Analytical expressions for the amplitudes for the Goldstino approximation}\label{appendixB}
In this appendix we present  explicitly the full results for  the 7 $\text{w}_{\psi_a\psi_a}^G$ functions that arose from 
the squared amplitudes [\ref{eq:amplitudesgoldstino}], as well as the 8 $\text{w}_{\psi_a\psi_b}^G$ functions that appear in the interference terms [\ref{eq:73:goldstino:interferences}] of the 3-body stop $\tilde{t}_1$ decay with goldstino in the final state.  First, we shall present the contribution for the squared amplitudes, then we shall present the interferences.  We shall shown that the  $\text{w}_{\psi_a\psi_a}^G$ and $\text{w}_{\psi_a\psi_b}^G$ functions are very compacts expressions, opposed to the resulting functions in the gravitino case that we have presented in Appendix  \ref{apendixA}. The approximation of the gravitino field by the derivative of the goldstino field  is good  in the high energy limit ($m_{\widetilde{G}}\ll m_{\tilde{t}_1}$), in the sense that in this limit they behave similar and also in the simplification of the computations.
\subsection{Top Contribution}
For the averaged squared amplitude of the top quark contribution, 
the resulting functions $\widetilde{\text{w}}_{jtt}$  $\forall\,j=1,2,3$ are:
\begin{align}\nonumber
\widetilde{\text{w}}_{1tt}&=4\frac{2 a_1 }{m_W^2}\Bigl(f_2 (m_W^2 (6
   m_{\tilde{G}}^2+2h_5 m_{\tilde{t}_1}^2-q_1^{2})+6
   f_3 m_W^2+4 f_3^2)-2 f_1
   (-m_{\tilde{G}}^2 (4 f_3+3
   m_W^2) \\\nonumber
   &\quad+f_2 (4 f_3+3
   m_W^2)+f_3 q_1^{2})+(q_1^{2}-2
   m_{\tilde{G}}^2) (m_W^2
   m_{\tilde{G}}^2+3 f_3 m_W^2+2
   f_3^2)\\ 
   &\quad+4 f_1^2
   (f_2-m_{\tilde{G}}^2)-2 m_W^2
   m_{\tilde{G}}^2h_5 m_{\tilde{t}_1}^2-4
  f_2^2 m_W^2\Bigr),\label{eq:36:w31:top-top.goldstino}
   \\ 
   \widetilde{\text{w}}_{2tt}&=\frac{4 a_2 m_{\tilde{G}}}{m_W^2}\bigl(m_W^2
   (m_{\tilde{G}}^2+m_{\tilde{t}}^2-2
   f_2)-f_1 (4 f_3+3 m_W^2)+3 f_3
   m_W^2+2 f_1^2+2
   f_3^2\bigr),\label{eq:37:w32:top-top.goldstino}
   \\ 
   \widetilde{\text{w}}_{3tt}&=\frac{2 a_3 (m_W^2
   (f_2-m_{\tilde{G}}^2)-3 f_3 m_W^2-2
   f_3^2+2 f_1 f_3)}{m_W^2}.
   \label{eq:38:w33:top-top.goldstino}
   \end{align}
With $a_1,\,a_2$ and $a_3$ defined previously in Appendix \ref{apendixA}.
\subsection{Sbottom Contribution}
For the averaged squared amplitude of the sbottom squark contribution, 
with the $\widetilde{\text{w}}_{1\tilde{b}_i\tilde{b}_i}$ function as:
\begin{equation}\label{eq:120}
\widetilde{\text{w}}_{1\tilde{b}_i \tilde{b}_i}=\frac{4 D_{\text{ij1}}h_2 h_3}{m_W^2},
\end{equation}
with $D_{ij1}$ defined previously in Appendix \ref{apendixA}.
\subsection{Chargino Contribution}
For the averaged squared amplitude of the chargino  contribution, 
the resulting functions $\widetilde{\text{w}}_{j\chi_i^+\chi_i^+}$  $\forall\,j=1,2,3$ are:
\begin{align}\nonumber
\widetilde{\text{w}}_{1\chi_i^+\chi_i^+} &=\frac{4 q_3^{2} P_{\text{ij1}}}{m_W^2}
 \Bigl(m_{\tilde{t}}^2
   (m_W^2 (m_{\tilde{G}}^2+f_2)+3 f_3
   m_W^2+2 f_3^2-2 f_1 f_3) \\
   &\quad +2 f_2
   (2 f_1 (f_1-f_3)-(3
   f_1+f_2) m_W^2)\Bigr),
   \label{eq:40:w35:chargino-chargino.goldstino}
   \\ 
   \widetilde{\text{w}}_{2\chi_i^+\chi_i^+} &=12m_{\tilde{G}}h_5 m_{\tilde{t}_1}^2 (P_{\text{ij1}}+P_{\text{ij2}}) 
   \left(h_5-f_1-f_2\right),\label{eq:41:w36:chargino-chargino.goldstino}
 \\ 
    \widetilde{\text{w}}_{3\chi_i^+\chi_i^+} &=\frac{4 P_{\text{ij2}}}{m_W^2} \Bigl(m_{\tilde{t}}^2
   (m_W^2 (m_{\tilde{G}}^2+f_2)+3 f_3
   m_W^2+2 f_3^2-2 f_1 f_3)\nonumber  \\
   &\quad +2 f_2
   (2 f_1 (f_1-f_3)-(3
   f_1+f_2) m_W^2)\Bigr),
   \label{eq:43:w38:chargino-chargino.goldstino}
\end{align}
where $P_{ij1}$ and $P_{ij2}$ are defined above in Appendix \ref{apendixA}.
\subsection{Interference Terms }
\subsubsection*{$\mathcal{M}_t^{G\,\dagger}\mathcal{M}_{\widetilde{b}_i}^G$ Interference}
For the interference term $\mathcal{M}_t^{G\,\dagger}\mathcal{M}_{\widetilde{b}_i}^G$, 
the functions $\text{w}_{jt\tilde{b}_i}$  $\forall\,j=1,2$ are:
\begin{align}\nonumber
\text{w}_{1t\tilde{b}_i}&=\frac{2 (\Delta _{\text{i1}}+\Delta
   _{\text{i2}})}{m_W^2} \Bigl(-f_1 (f_3
   (m_{\tilde{t}}^2-m_{\tilde{G}}^2)+f_2
   m_W^2)+m_W^2 \bigl(m_{\tilde{t}}^2 (2
   m_{\tilde{G}}^2+f_3) \\
   &\quad-f_2
   (m_{\tilde{G}}^2+m_{\tilde{t}}^2)\bigr)
   +2 f_1^2
   (f_2-m_{\tilde{G}}^2)\Bigr),
   \label{eq:45:w39:top-sbotom.goldstino}
   \\ 
   \text{w}_{2t\tilde{b}_i} &= \frac{2 m_{\tilde{G}} (\Delta _{\text{i1}}-\Delta
   _{\text{i2}}) (m_W^2
   (f_2-m_{\tilde{t}}^2)+f_1^2-f_3 f_1)}{m_W^2}.
   \label{eq:46:w40:top-sbotom.goldstino}
\end{align}
Where $\Delta _{\text{i1}}$ and $\Delta _{\text{i2}}$ are defined above in Appendix \ref{apendixA}.
\subsubsection*{$\mathcal{M}_{\chi_i^+}^{G\,\dagger}\mathcal{M}_{\widetilde{b}_i}^G$ Interference}
For the interference term $\mathcal{M}_{\chi_i^+}^{G\,\dagger}\mathcal{M}_{\widetilde{b}_i}^G$,
the functions  $\text{w}_{j\chi_i^+\tilde{b}_i}^G$ $\forall\,j=1,2$ are:
\begin{align}
\text{w}_{1\chi_i^+\tilde{b}_i}^G &=\frac{4 C_{\text{ij1}} m_{\tilde{G}}}{m_W^2}
   \left(m_{\tilde{t}}^2 (m_W^2
   (h_5-f_2)+f_1
   f_3)-f_1^2
   h_5\right),
   \label{eq:52:w45:chargino-sbotom.goldstino}
   \\ 
  \text{w}_{2\chi_i^+\tilde{b}_i}^G&=\frac{4 C_{\text{ij2}}}{m_W^2}\bigl(-m_W^2h_5 m_{\tilde{t}_1}^2
   h_2+f_1^2 (2 f_2-m_{\tilde{G}}^2)-f_1
   (f_3h_5 m_{\tilde{t}_1}^2+f_2
   m_W^2)\bigr),
   \label{eq:53:w46:chargino-sbotom.goldstino}
\end{align}
with $ C_{\text{ij1}}$ and $C_{\text{ij2}}$ defined above in Appendix \ref{apendixA}.

\subsubsection*{$\mathcal{M}_t^{G\dagger}\mathcal{M}_{\chi_i^+}^G$ Interference}
For the interference term $\mathcal{M}_t^{G\dagger}\mathcal{M}_{\chi_i^+}^G$,
the functions $\text{w}_{jt\chi_i^+}^G$  $\forall\,j=1,2,3,4$ are:
\begin{align}\nonumber
\text{w}^G_{1t \chi_i^+}&=\frac{4 R_{\text{i1}}\, m_{\tilde{G}}}{m_W^2}
   \Bigl(m_{\tilde{t}}^2 (m_W^2 (4
   m_{\tilde{G}}^2-3 f_2+3 m_W^2)+5 f_3 m_W^2-2
   f_3^2) \\  \nonumber
   &\quad+f_2 m_W^2 (2 f_2-3
  h_5)-2
   f_1^2 m_{\tilde{G}}^2+f_1 (4
   f_2 (f_3+m_W^2)-3 m_W^2
  h_5 m_{\tilde{t}_1}^2)\Bigr),
   \\ 
 \text{w}_{2t \chi_i^+}^G&=\frac{4 R_{\text{i2}}}{m_W^2} \Bigl(m_{\tilde{t}}^2 (f_3
   (2 f_3+m_W^2)-m_W^2
   m_{\tilde{G}}^2)+f_2 m_W^2 (3
  h_5
   -2
   f_2)\nonumber  \\ 
  &\quad  +2 f_1^2 m_{\tilde{G}}^2-4
   f_1 f_2 (f_3+m_W^2)\Bigr),
\\ 
\text{w}_{3t \chi_i^+}^G&=\frac{4 R_{\text{i3}}\, m_{\tilde{G}}}{m_W^2} \left(m_W^2
   (m_{\tilde{t}}^2-f_2)-f_1 (2 f_3+3
   m_W^2)+2 f_1^2\right),
\\ 
\text{w}_{4t \chi_i^+}^G&=\frac{R_{\text{i4}}}{m_W^2} \Bigl(4 (2 f_1
   (f_1-f_3)-(3 f_1+f_2)
   m_W^2) (2 f_2-m_{\tilde{G}}^2) \\ \nonumber
  &\quad +4
  h_5 m_{\tilde{t}_1}^2 ((f_2+3 f_3)
   m_W^2+2 f_3
   (f_3-f_1))\Bigr),
\end{align}
with $R_{\text{i1}},\,R_{\text{i2}},\,R_{\text{i3}}$ and $R_{\text{i4}}$  defined above in Appendix \ref{apendixA}.
\section*{Acknowledgments}
We would like to acknowledge the support of CONACYT and SNI. We also acnknowledge to Abdel Perez for his valuable comments. B.O. Larios is supported by a CONACYT graduate student
fellowship.

\end{document}